\documentclass[aps,pra,twocolumn,showpacs,floatfix,superscriptaddress]{revtex4}
\usepackage{graphicx}
\usepackage{dcolumn}
\usepackage{amsmath}
\usepackage{amssymb}
\usepackage{dsfont}
\usepackage{bm}
\usepackage{epstopdf}
\newcommand{\fig}[1]{Figure~\ref{#1}}

\begin{document}

\title{Recurrences in three-state quantum walks on a plane}
\author{B. Koll\'ar}
\affiliation{Department of Quantum Optics and Quantum Information, Research
Institute for Solid State Physics and Optics, Hungarian Academy of
Sciences, Konkoly-Thege Miklós út 29-33, H-1121 Budapest, Hungary}

\author{M. \v Stefa\v n\'ak\email[correspondence to:]{martin.stefanak@fjfi.cvut.cz}}
\affiliation{Department of Physics, Faculty for Nuclear Sciences and Physical Engineering, Czech Technical University in Prague, B\v
rehov\'a 7, 115 19 Praha 1 - Star\'e M\v{e}sto, Czech Republic}

\author{T. Kiss}
\affiliation{Department of Quantum Optics and Quantum Information, Research
Institute for Solid State Physics and Optics, Hungarian Academy of
Sciences, Konkoly-Thege Miklós út 29-33, H-1121 Budapest, Hungary}

\author{I. Jex}
\affiliation{Department of Physics, Faculty for Nuclear Sciences and Physical Engineering, Czech Technical University in Prague, B\v
rehov\'a 7, 115 19 Praha 1 - Star\'e M\v{e}sto, Czech Republic}


\date{\today}

\begin{abstract}
We analyze the role of dimensionality in the time evolution of
discrete time quantum walks through the example of the three-state
walk on a two-dimensional, triangular lattice. We show that the
three-state Grover walk does not lead to trapping (localization) or
recurrence to the origin, in sharp contrast to the Grover walk on
the two dimensional square lattice. We determine the power law
scaling of the probability at the origin with the method of
stationary phase. We prove that only a special subclass of coin
operators can lead to recurrence and there are no coins leading to
localization. The propagation for the recurrent subclass of coins is
quasi one-dimensional.
\end{abstract}

\maketitle

%%%%%%%%%%%%%%%%%%%%%%%%%%%%%%%%%%%%%%%%%%%%%%%%%%%%%%%%%%%%%%%

\section{Introduction}

The speed of propagation is affected critically by the spatial
dimensions in diverse systems, ranging from classical statistical
phenomena like diffusion or percolation \cite{percolation,odor} to
various quantum transport effects \cite{nazarov}. One of the
simplest models capturing the essence of classical diffusive
propagation is a random walk on a regular lattice \cite{revesz}. The
sensitivity of a random walk to the number of spatial dimensions is
clearly shown by its recurrence properties. A balanced
random walker on a regular square lattice returns to its
starting point with certainty in dimensions 1 and 2 if we wait long
enough, whereas there is a nonzero probability of escape in
dimension 3 or higher, as discovered by P\'olya \cite{polya}.
The probability to return to the origin is now called the
P\'olya number.

The generalization of the discrete time random walk to a quantum
walk was first introduced by Aharonov, Davidovich and Zagury
\cite{aharonov_1993}, and independently, from a quantum information
theoretical perspective, by Meyer \cite{meyer}. In the past few
years, there has been a considerable number of studies devoted to
quantum walks \cite{santha_2008}, strongly motivated by their
possible application for designing quantum algorithms
\cite{kendon_2006}.  In  discrete time walks the external, position
degrees of freedom are  assisted by a finite number of internal
degrees of freedom, the chirality or coin. The situation resembles
to the discrete version of the Chalker-Coddington model
\cite{chalker-coddington}. Recent experimental demonstration of
discrete-time quantum walks with many steps in different physical
systems (trapped ions \cite{zahringer}, optically trapped atoms
\cite{karski}, linear quantum optics \cite{schreiber_2010}) pave the
way to study more complex behavior of quantum walks in the
laboratory.

In quantum walks, the question of finding the walker involves the
measurement of its position. The measurement unavoidably disturbs
the quantum system, which has to be taken into account when defining
quantities to describe its spreading, like the hitting time
\cite{hittingreview}. We have defined the generalized P\'olya number
\cite{stefanak_prl_2008} to characterize the probability of return
to the origin of a quantum walker starting from a localized
position. In order to minimize the disturbance we have adopted a
scheme for measurements, where each system from an ensemble is
observed only once \cite{kiss-kecskes}. The quantum P\'olya number
exhibits strikingly different behaviour compared to its classical
counterpart. In dimension 2, the symmetric quantum walk on a square
lattice can be recurrent (P\'olya number 1) or transient (P\'olya
number less than 1) depending on the coin operator and the initial
state \cite{stefanak_pra_2008}. For certain coins,  localization
(trapping) at the origin can occur \cite{localization}. In dimension
1, although both classical and quantum unbiased walks have a P\'olya
number 1, classically the effect is unstable towards the
introduction of a bias while in the quantum case there is a region
of stability \cite{stefanak_njp_2009}. These results indicate that
the recurrence of walks is not only sensitive to the dimension of
the underlying lattice, but in the quantum case also the coin degree
of freedom can ultimately determine its behaviour. In the one
dimensional case, one can increase the dimension of the coin space
from 2 to 3, by allowing the walker to stay at its position leading
to localization \cite{inui_2005}. Further increasing the dimension
of the coin space can possibly introduce more interesting effects
\cite{inui_2003}.

The triangular lattice is a planar graph with symmetry properties different from the square lattice. It can define either a rank 3 oriented graph or a rank 6 undirected graph. Quantum walks on triangular lattices have been considered for designing effective quantum algorithms  \cite{abal_arxiv_2010}. Discrete-time quantum walks, especially on triangular lattices, provide a platform to realize topological phases \cite{kitagawa}. Propagation of continuous-time quantum walks have also been recently studied in \cite{jafarizadeh}. 

In this paper, we consider a two dimensional walk on an oriented triangular graph with a three-state coin space. The classical symmetric random walk on such a graph is recurrent, similar to the two-dimensional square lattice. We analyze the quantum walk on this lattice and prove that neither localization nor recurrence is possible for any unbiased coin, except for trivial cases. We calculate the asymptotic decay of the probability at the origin for the Grover coin and discuss its dependence on the initial state.

The paper is  organized as follows. In Section \ref{desc}. we introduce the three-state quantum walk on the triangular lattice and  describe the asymptotic behavior of such a quantum walk with the help of the method of stationary phase. In Section \ref{secgrowalk}. we analyze the recurrence properties of the walk driven by the Grover matrix. We dedicate Section \ref{recthreestate}. to find the general requirements of the recurrence in arbitrary coin matrices, we prove that only a special subclass of coins leads to recurrence, we demonstrate the quasi-onedimensional propagation leading to recurrence of such walks by constructing an example. We summarize our results in Section \ref{secsummary}. For the sake of completeness we show the recurrence properties of the classical walk on the triangular lattice in Appendix \ref{app:a}.

\section{Description of the three-state quantum walks}
\label{desc}

\begin{figure}
\includegraphics[width=0.50\textwidth]{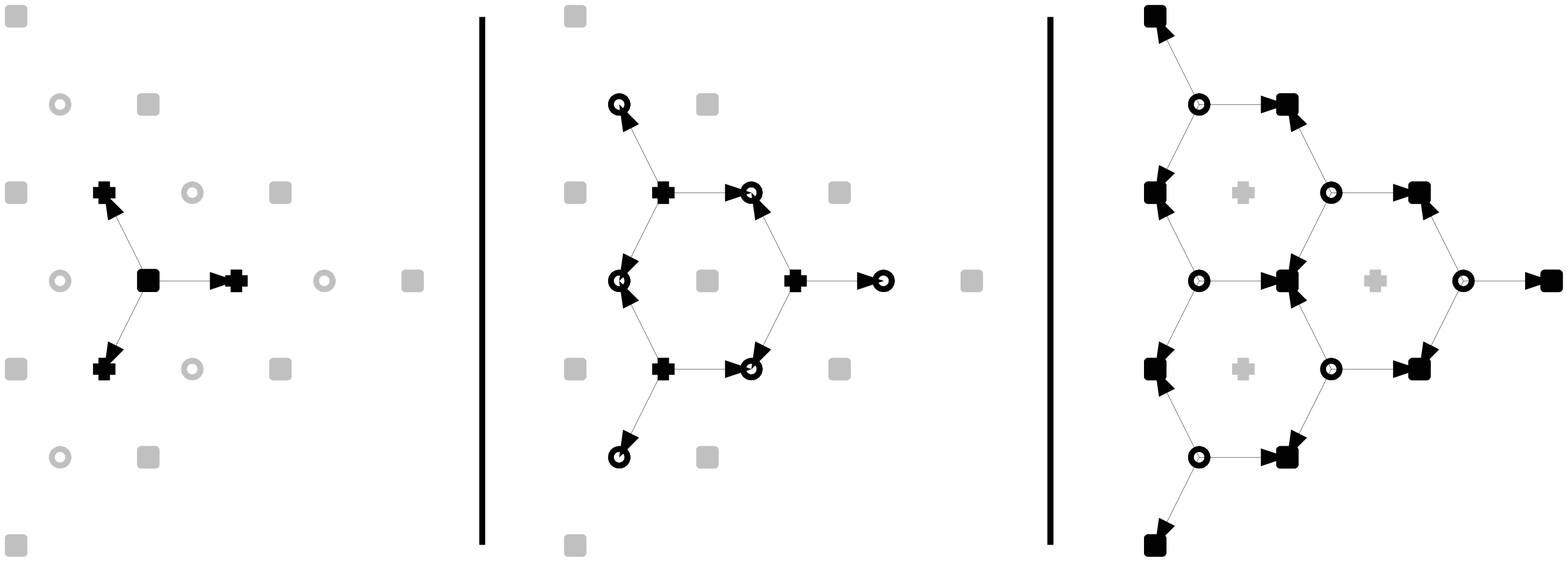}
\caption{The lattice points which the walker starting from the origin (central vertex) can reach in 3 steps, with the available displacement vectors.}
\label{fig:topology}
\end{figure}

The Hilbert space of the three-state quantum walk on the $2$-dimensional lattice has the tensor product structure
\begin{equation}
\label{def:hilbert}
\mathcal{H}=\mathcal{H}_P \otimes \mathcal{H}_C\,,
\end{equation}
where $\mathcal{H}_P$ denotes the so-called position-space spanned by the vectors $|\textbf{m}\rangle$ corresponding to the walker being at the lattice point $\textbf{m}$
\begin{equation}
\label{def:pspace}
\mathcal{H}_P= {\rm Span} \left\{|\textbf{m}\rangle \, | \,\, \textbf{m}\in\mathds{Z}^2\right\} = \ell^2(\mathds{C}^2)\,.
\end{equation}
The three-dimensional coin-space $\mathcal{H}_C$ has the structure
\begin{equation}
\label{def:cspace}
\mathcal{H}_C=\text{Span}\left\{|\mathbf{e}_i\rangle\,|\,\, i=1, 2, 3\right\}\,,
\end{equation}
where the coin state (chirality) $|\mathbf{e}_i\rangle$ corresponds to the displacement of the particle by $\mathbf{e}_i$, a two dimensional vector on the plane of walk. The triangle-like topology of the walk (see \fig{fig:topology}) leads to the following form of the displacement vectors
\begin{equation}
\mathbf{e}_1=(-\frac{1}{2}, \frac{\sqrt{3}}{2}),\quad\mathbf{e}_2=(1, 0),\quad\mathbf{e}_3=(-\frac{1}{2}, -\frac{\sqrt{3}}{2})\,.
\end{equation}
The walks in consideration are unbiased in a sense, that
\begin{equation}
\label{average:disp}
\sum_{i=1}^3{\mathbf{e}_i}=0\,.
\end{equation}

A single step of the quantum walk is given by the propagator
\begin{equation}
\label{time:evolution}
U=S \cdot (I_P \otimes C)\,.
\end{equation}
Here $I_P$ denotes the identity operator on $\mathcal{H}_P$. The operator $C$ represents the coin flip and acts only on $\mathcal{H}_C$. The conditional step operator $S$ is responsible for the actual displacement of the walker from position $\textbf{m}$ to $\textbf{m}+\mathbf{e}_i$ with respect to the coin state $|\mathbf{e}_i \rangle$ and has the following form
\begin{equation}
\label{def:step}
S=\sum_{\mathbf{m}} \sum_{i=1}^3{ | \textbf{m} + \mathbf{e}_i\rangle \langle \textbf{m}| \otimes | \mathbf{e}_i \rangle \langle \mathbf{e}_i |}\,.
\end{equation}

The state of the walker after $t$ time steps is given by the application of the time evolution operator (\ref{time:evolution}) on the initial state
\begin{equation}
\label{def:psievol}
|\psi(t)\rangle \equiv \sum_{\textbf{m},i} \psi_i(\textbf{m},t)|\mathbf{m}\rangle \otimes |\mathbf{e}_i \rangle = U^t |\psi(0)\rangle\,.
\end{equation}
The probability of finding the walker at the lattice point $\textbf{m}$ at time $t$ is given by the summation over the coin states
\begin{eqnarray}
\label{def:prob}
p(\textbf{m},t) & \equiv & \sum_{i=1}^3 | \langle \textbf{m} | \otimes \langle \mathbf{e}_i | \psi(\textbf{m},t) \rangle |^2 = \sum_{i=1}^3 |\psi(\textbf{m},t)|^2 \nonumber \\
 & = & ||\psi(\textbf{m},t)||^2,
\end{eqnarray}
Here, we have introduced the vector of probability amplitudes at the lattice point $\mathbf{m}$
\begin{equation}
\psi(\textbf{m},t) \equiv (\psi_1(\textbf{m},t),\psi_2(\textbf{m},t),\psi_3(\textbf{m},t))^T\,.
\end{equation}

Since we focus on the recurrence properties of the three-state quantum walks we consider initial states according to the classical P\'olya problem, i.e. the walker starts localized at the origin. Hence, the probability amplitudes vanishes except at the origin
\begin{equation}
\psi_i(\textbf{m},0) = 0, \quad \mathbf{m}\neq \mathbf{0}\,.
\end{equation}
Nevertheless, we have the freedom to choose the initial orientation of the coin which we denote by the vector of probability amplitudes $\psi = \left(\psi_1(\textbf{0},0),\psi_2(\textbf{0},0),\psi_3(\textbf{0},0)\right)^T.$

Since the three-state quantum walk is translationally invariant the time evolution equation (\ref{def:psievol}) is greatly simplified by the Fourier transformation
\begin{equation}
\label{fourier:transform}
\tilde{\psi}(\textbf{k},t) \equiv \sum_{\textbf{m}} \psi(\textbf{m},t) e^{i \textbf{k} \cdot \textbf{m}}, \qquad \textbf{k} \in \mathds{K}^2\,,
\end{equation}
where $\mathds{K}=(-\pi,\pi]$. The time evolution in the Fourier picture simplifies into a single difference equation
\begin{equation}
\label{fourier:evol}
\tilde{\psi}(\textbf{k},t) = \tilde{U}^t(\textbf{k}) \tilde{\psi}(\textbf{k},0) =  \tilde{U}^t(\textbf{k}) \psi \,.
\end{equation}
where $\tilde{\psi}(\textbf{k},0)$ denotes the Fourier transformation of the initial state. The propagator in the Fourier picture $\tilde{U}$ is given by
\begin{equation}
\label{fourier:def:u}
\tilde{U} = D(\textbf{k}) \cdot C\,,
\end{equation}
where $D(\textbf{k})$ is a diagonal matrix determined by the displacement vectors $\mathbf{e}_i$
\begin{equation}
\label{fourier:def:d}
D(\textbf{k}) = \mathrm{Diag}\left( e^{-i \textbf{k} \cdot \mathbf{e}_1}, e^{-i \textbf{k} \cdot \mathbf{e}_2}, e^{-i \textbf{k} \cdot \mathbf{e}_3} \right)\,.
\end{equation}

The time evolution equation in the Fourier picture (\ref{fourier:evol}) is readily solved by diagonalizing the propagator $\tilde{U}$. Since $\tilde{U}$ is an unitary matrix its eigenvalues have the form
\begin{equation}
\label{fourier:u:eigenvalues}
\lambda_j(\textbf{k}) = e^{i
\omega_j(\textbf{k})}\,.
\end{equation}
The solution in the position representation reads
\begin{equation}
\label{fourier:inverse}
\psi(\textbf{m},t) = \sum_{j = 1}^3 \int_{\mathds{K}^2} \frac{d \textbf{k}}{(2 \pi)^2} e^{- i\textbf{k}\cdot\textbf{m}} e^{i \omega_j(\textbf{k}) t} \left( v_j(\textbf{k}),\psi\right) v_j(\mathbf{k})\,.
\end{equation}
We concentrate on the recurrence nature of the quantum walks which is determined by the asymptotic behaviour of the probability at the origin. This can be readily analyzed by means of the method of stationary phase. Indeed, the amplitude at the origin reads
\begin{equation}
\label{psi:0}
\psi(\textbf{0},t) = \sum_{j = 1}^3 \int_{\mathds{K}^2} \frac{d \textbf{k}}{(2 \pi)^2} e^{i \omega_j(\textbf{k}) t} \left( v_j(\textbf{k}),\psi\right) v_j(\mathbf{k})\,,
\end{equation}
and the probability is simply the absolute square of the amplitude.
Within the stationary phase approximation the important points in
the integration domain are those where the phase
$\omega_j(\textbf{k})$ has a vanishing derivative, i.e. stationary
points. The rate at which the integrals in (\ref{psi:0}) decay is
determined by the flatness of the phase at the stationary points.
For a two-dimensional walk with a finite number of
non-degenerate saddle points the probability amplitude at the
origin decays as $t^{-1}$, with the probability decaying as
$t^{-2}$ leading to a transient walk. A continuum of saddle
points (saddle line) leads to a probability amplitude
decaying at a rate $t^{-1/2}$, the probability decays
as $t^{-1}$ at the origin, thus resulting in recurrence.

%%%%%%%%%%%%%%%%%%%%%%%%%%%%%%%%%%%%%%%%%%%%%%%%%%%%%%%%%%%%%%%%%%%%%%%%%%%%%%%

\section{Grover walk}
\label{secgrowalk}

The Grover operator plays a key role in Grover's search algorithm.
Used as a coin operator for regular two-dimensional quantum walks on
a Cartesian lattice it leads to the phenomenon of localization for
all initial coin states except one well defined state. Moreover it
is widely used in quantum walk based search algorithms.

The Grover matrix is an orthogonal matrix with elements
defined as
\begin{equation}
G^{(d)}_{i,j} = \frac{2}{d} - \delta_{i,j}\,.
\end{equation}

This matrix has an important symmetry. Indeed, it commutes with all
permutation matrices. Hence, in the $d=3$ case cyclic permutation of
the initial chiralities will only rotate the probability
distribution by $2 \pi/3$ in the positive or the negative direction.
To obtain a rotationally invariant probability distribution we have
to choose an initial state which is invariant under cyclic
permutations. Since the global phase of the quantum state is
irrelevant we find such a symmetric probability distribution results
from the initial coin state
\begin{equation}
\psi_S = \frac{1}{\sqrt{3}}\left(1,1,1\right)^T\,.
\label{grover:sym}
\end{equation}
Moreover, the symmetry of the Grover operator implies that permuting the initial chiralities does not change the recurrence properties of the Grover walk.

Let us analyze the recurrence of the three-state Grover walk in more
detail. For that purpose we have to find the asymptotic behaviour of
the probability at the origin. This is determined by the stationary
points of the eigenenergies $\omega_j(\mathbf{k})$ of the propagator
$\tilde{U}(\mathbf{k})$ in the Fourier picture. For the Grover walk
the eigenenergies are given by the implicit function
\begin{eqnarray}
\nonumber \Phi(\omega,\mathbf{k}) & = & \det\left(\tilde{U}(\mathbf{k}) - e^{i\omega} I\right) = \\
\nonumber & = & \sin \left(k_1 - \frac{\omega}{2}\right) -  2 \cos \left(  \frac{\sqrt{3} k_2}{2} \right) \sin \left( \frac{k_1 + \omega}{2} \right) - \\
\nonumber & - &  3 \sin\left(\frac{3\omega}{2}\right) = 0\,.\\
\end{eqnarray}
It is straightforward to show by implicit differentiation that the stationary point is $\mathbf{k}_0=\mathbf{0}$. Moreover, also the second derivatives are vanishing at $\mathbf{k}_0$. To clarify this statement, we consider the eigenvalues of $\tilde{U}(\mathbf{k})$ for $k_1=0$ or $k_2= \frac{1}{\sqrt{3}} k_1$. In both cases we find that one eigenvalue, say $\lambda_1$, is constant and equals unity, i.e.
\begin{equation}
\omega_1(k_1 = 0,k_2) = \omega_1(k_1,k_2 = \frac{1}{\sqrt{3}} k_1) = 0\,.
\end{equation}
Hence, all derivatives of $\omega_1$ with respect to $k_1$ at the line $k_2 =  \frac{1}{\sqrt{3}} k_1$ vanish, the same applies on the line $k_1 = 0$ with taking the derivatives with respect to $k_2$. Therefore, the second derivatives of $\omega_1(\mathbf{k})$ vanish at the stationary point $\mathbf{k}_0 = \mathbf{0}$, i.e. the rank of the Hessian matrix at $\mathbf{k}$ is zero. In such a case, the method of stationary phase indicates that the integrals in the probability amplitude (\ref{psi:0}) decay like $t^{-2/3}$ as $t$ approaches infinity. Hence, the decay of the probability at the origin is given by $p_0(t)\sim t^{-4/3}$ which is faster than the threshold required for the recurrence. We conclude that the three-state Grover walk on a plane is transient, i.e. its P\'olya number is less than unity.

The decay of the probability at the origin can be even faster than what we have already found. Indeed, we can eliminate the stationary point $\mathbf{k}_0$ by the proper choice of the initial coin state $\psi$. Such $\psi$ has to be orthogonal to the eigenvector corresponding to $\lambda_1$ at the stationary point $\mathbf{k}_0$, which is easily found to be
\begin{equation}
v_1(\mathbf{k}_0) = \frac{1}{\sqrt{3}}\left(1,1,1\right)^T\,.
\end{equation}
Hence, the initial coin state has to be of the form
\begin{equation}
\psi_G  = \left(a,b,c\right)^T,\quad a+b+c = 0\,.
\label{psi:fast}
\end{equation}
Starting the three-state Grover walk with an initial state from the subspace (\ref{psi:fast}) will lead to fast decay of the probability at the origin.

\begin{figure}
\includegraphics[width=0.4\textwidth]{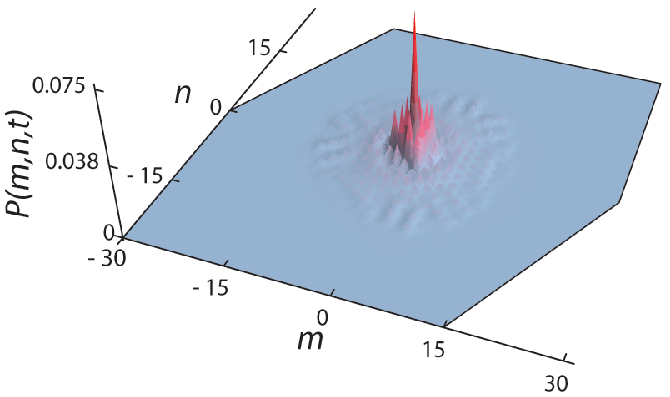}\vspace{12pt}
\includegraphics[width=0.4\textwidth]{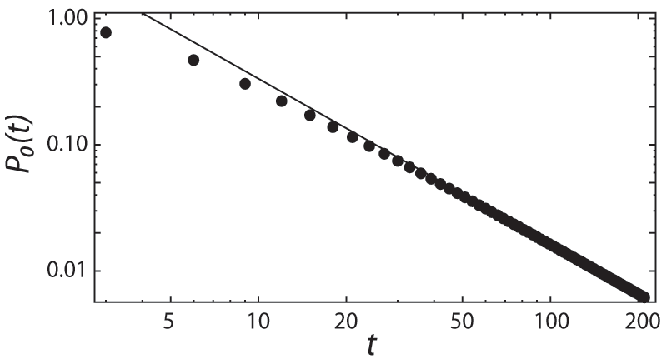}
\caption{(Color online) Probability distribution of the three-state Grover walk on
a plane and the time dependence of the probability at the origin (throughout the paper time $t$ is measured in dimensionless units, corresponding to one step). In
the upper plot we display the probability distribution after 30
steps. The initial coin state was chosen to be $\psi_S =
\frac{1}{\sqrt{3}}\left(1,1,1\right)^T$. We find that the
probability distribution is symmetric and peaked at the origin.
However, the probability at the origin is decaying according to a
power law. The lower plot is in loglog scale to unravel the scaling
law on a longer time scale. The straight line corresponds to
$t^{-4/3}$.} \label{fig:grover1}
\end{figure}

\begin{figure}
\includegraphics[width=0.4\textwidth]{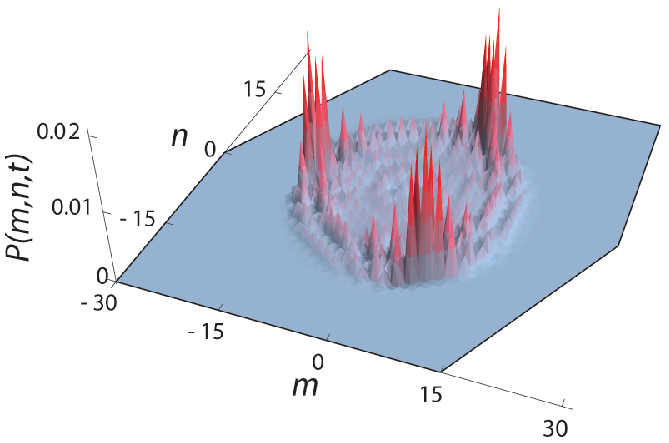}\vspace{12pt}
\includegraphics[width=0.4\textwidth]{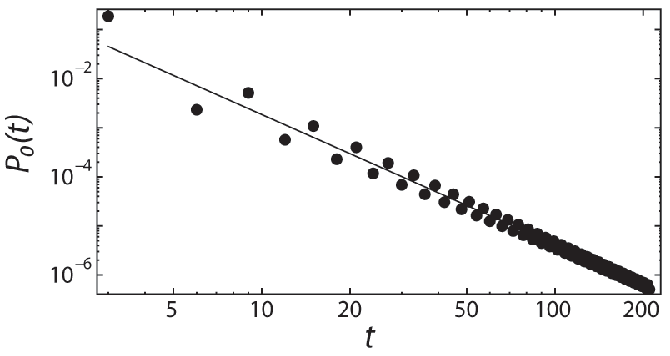}
\caption{(Color online) Probability distribution of the three-state Grover walk on
a plane and the time dependence of the probability at the origin for
the initial state $\psi = \frac{1}{\sqrt{3}}\left(1,\exp({i
2\pi/3}),\exp({i 4\pi/3})\right)^T$. Since $\psi$ belongs to the
family (\ref{psi:fast}) the behaviour of the Grover walk is changed
considerably. The peak at the origin has vanished, as illustrated in
the upper plot. Most of the probability is located on the
ring. Moreover, the decay of the probability at the origin is
faster when compared to the power law we have found for the initial
states outside of the subspace (\ref{psi:fast}) . On the lower plot
we show the long time behaviour of the probability at the origin,
thus indicating that the exponent of the power law has doubled to
$-8/3$. We use a loglog scale to make the power law more
visible. The straight line corresponds to $t^{-8/3}$.}
\label{fig:grover2}
\end{figure}

We illustrate these results in Figures~\ref{fig:grover1} and \ref{fig:grover2}. In Figure~\ref{fig:grover1} we choose the initial state $\psi_S$ of (\ref{grover:sym}). The resulting probability distribution shown in the upper plot is symmetric and peaked at the origin. However, in contrast to the Grover walk on a Cartesian lattice, the current model does not exhibit localization. In the lower plot, $p_0(t)$ is shown on a double logarithmic scale. The numerical results agree with the power law $p_0(t) \sim t^{-4/3}$ predicted by the method of stationary phase.

In Figure~\ref{fig:grover2}, the initial state was chosen to be $\psi = \frac{1}{\sqrt{3}}\left(1,e^{i 2\pi/3},e^{i 4\pi/3}\right)^T$ which belongs to the subspace (\ref{psi:fast}). The peak of the probability distribution at the origin vanishes (upper plot). The numerical results of the probability at the origin presented in the lower plot indicate that the decay rate has doubled to $p_0(t) \sim t^{-8/3}$.

%%%%%%%%%%%%%%%%%%%%%%%%%%%%%%%%%%%%%%%%%%%%%%%%%%%%%

\section{Recurrence in the three-state quantum walk}
\label{recthreestate}

After analyzing a particular case we turn our attention to the
recurrence behaviour of a general three state quantum walk. Let us
consider an arbitrary ${\rm SU}(3)$ coin matrix $C$ with matrix
elements $C_{ij}$. The characteristic polynomial of the propagator
in the Fourier picture $\tilde{U}(\mathbf{k})$ has the form
\begin{equation}
\label{charpoly:unitary}
\lambda^3 - \lambda^2{\rm Tr}\tilde{U}(\mathbf{k})
 + \lambda\left[\Delta_{12}(\mathbf{k}) + \Delta_{13}(\mathbf{k}) + \Delta_{23}(\mathbf{k})\right] - 1 = 0\,,
\end{equation}
where we denote
\begin{equation}
\Delta_{ij}(\mathbf{k}) = e^{-i \textbf{k} \cdot (\mathbf{e}_i + \mathbf{e}_j)}\left(C_{ij}C_{ji} - C_{ii}C_{jj}\right)\,.
\end{equation}
We have also used the fact that the determinant of $\tilde{U}(\mathbf{k})$ is unity. With the help of the eigenvalues $\lambda_j = e^{i \omega_j(\textbf{k})}$ of the matrix $\tilde{U}(\mathbf{k})$ we can express the characteristic polynomial (\ref{charpoly:unitary}) in a different form
\begin{eqnarray}
\label{charpoly:eigen}
\nonumber & \lambda^3 & - \lambda^2 \left(e^{i \omega_1(\textbf{k})} + e^{i \omega_2(\textbf{k})} + e^{i \omega_3(\textbf{k})}\right) + \\
\nonumber & + & \lambda \left(e^{i (\omega_1(\textbf{k}) + \omega_2(\textbf{k}))} + e^{i (\omega_1(\textbf{k}) + \omega_3(\textbf{k}))} + e^{i (\omega_2(\textbf{k}) + \omega_3(\textbf{k}))}\right) - \\
& - & e^{i (\omega_1(\textbf{k}) + \omega_2(\textbf{k}) + \omega_3(\textbf{k}))} = 0 \,.
\end{eqnarray}
Comparing the coefficients of the same powers of $\lambda$ in the expressions (\ref{charpoly:unitary}) and (\ref{charpoly:eigen}) we find the relations
\begin{eqnarray}
\label{charpoly:det} (\lambda^0): & e^{i (\omega_1(\textbf{k}) + \omega_2(\textbf{k}) + \omega_3(\textbf{k}))} = 1 \\ \nonumber \\
\nonumber (\lambda^1): & e^{i (\omega_1(\textbf{k}) + \omega_2(\textbf{k}))} + e^{i (\omega_1(\textbf{k}) + \omega_3(\textbf{k}))} + e^{i (\omega_2(\textbf{k}) + \omega_3(\textbf{k}))} = \\
 & = \Delta_{12}(\mathbf{k}) + \Delta_{13}(\mathbf{k}) + \Delta_{23}(\mathbf{k}) \\ \nonumber \\
\label{charpoly:trace} (\lambda^2): & e^{i \omega_1(\textbf{k})}+e^{i \omega_2(\textbf{k})}+e^{i \omega_3(\textbf{k})} = {\rm Tr}\tilde{U}(\mathbf{k}) \,.
\end{eqnarray}
These equations must be satisfied for all $\textbf{k} \in \mathds{K}^2$.

From the relation (\ref{charpoly:det}) we could easily replace one of the eigenvalues with
\begin{equation}
\omega_3(\mathbf{k}) = -\omega_1(\mathbf{k}) -\omega_2(\mathbf{k})\,.
\end{equation}
With this result the equation (\ref{charpoly:trace}) simplifies into
\begin{eqnarray}
\label{charpoly:trace:simp}
e^{i \omega_1(\mathbf{k})} + e^{i \omega_2(\mathbf{k})} + e^{-i (\omega_1(\mathbf{k}) + \omega_2(\mathbf{k}))} =  {\rm Tr}\tilde{U}(\mathbf{k}) =  \nonumber\\ =e^{-i \textbf{k} \cdot \mathbf{e}_1} C_{11} +  e^{-i \textbf{k} \cdot \mathbf{e}_2 } C_{22}  +  e^{-i \textbf{k} \cdot \mathbf{e}_3} C_{33} \,, \nonumber \\
\end{eqnarray}
where we have expressed the trace of the propagator $\tilde{U}(\mathbf{k})$ explicitly. By multiplying this equation with $e^{i \frac{1}{2} \omega_1(\mathbf{k})}$ and taking the imaginary part we find
\begin{eqnarray}
\label{charpoly:trace:imsimp}
&& \sin\left(\frac{3}{2}  \omega_1(\mathbf{k})  \right)  = \nonumber \\
& = & |C_{11}| \sin \left( -\textbf{k} \cdot \mathbf{e}_1 + \mathrm{Arg}\left( C_{11} \right) + \frac{1}{2}  \omega_1(\mathbf{k}) \right) + \nonumber\\
 & + & |C_{22}| \sin \left( -\textbf{k} \cdot \mathbf{e}_2 + \mathrm{Arg}\left( C_{22} \right) + \frac{1}{2}  \omega_1(\mathbf{k}) \right) + \nonumber \\
& + & |C_{33}| \sin \left( -\textbf{k} \cdot \mathbf{e}_3 + \mathrm{Arg}\left( C_{33} \right) + \frac{1}{2}  \omega_1(\mathbf{k}) \right)
\,.
\end{eqnarray}

We focus on the recurrence properties of the quantum walk, thus the
number of the saddle points ($\mathbf{k_0}$).  Let us assume that in
$\omega_1(\mathbf{k})$ we have a continuum of saddle points (a
"saddle line"). On this line ($\mathbf{k_0}$) the gradient of
$\omega_1(\mathbf{k})$ must vanish in respect of $\mathbf{k}$
\begin{equation}
\label{saddle:line:gradient:vanish}
\left(\frac{\partial \omega_1(\mathbf{k})}{\partial \mathbf{k}}\right)_{\mathbf{k}=\mathbf{k_0}} = \mathbf{0} \,.
\end{equation}
One could easily consider that on the saddle line the value of
$\omega_1(\mathbf{k_0})$ must be constant, thus we could substitute
$\omega_1(\mathbf{k_0})$ with a constant $2 \phi$. By taking the
gradient in respect of $\mathbf{k}$ of Eq.
(\ref{charpoly:trace:imsimp}) and using the previous assumptions
(moving to the saddle line $\mathbf{k}=\mathbf{k_0}$ and
substituting $\omega_1(\mathbf{k})$ with $2\phi$ and its gradient
with $\mathbf{0}$) we have
\begin{eqnarray}
\mathbf{0} & = & \mathbf{e}_1  |C_{11}| \cos \left( - \mathbf{k_0} \cdot \mathbf{e}_1  + \mathrm{Arg}\left( C_{11} \right) + \phi  \right) + \nonumber \\
 & + & \mathbf{e}_2  |C_{22}|  \cos \left(   - \mathbf{k_0} \cdot \mathbf{e}_2  + \mathrm{Arg}\left( C_{22} \right) + \phi  \right) + \nonumber \\
& + & \mathbf{e}_3  |C_{33}|  \cos \left(     -\mathbf{k_0} \cdot \mathbf{e}_3 + \mathrm{Arg}\left( C_{33} \right) + \phi  \right) \,.
\label{derived:saddle:line}
\end{eqnarray}
The last equation can be satisfied for a continuum of saddle points
$\mathbf{k_0}$ (saddle line) only if two of the absolute values
$|C_{11}|, |C_{22}|, |C_{33}|$ are zero. It is easy to prove that in
this case the walk is quasi $1$-dimensional. To show that let us
assume that $C_{11}$ and $C_{33}$ are both zero. The unitarity of
the coin operator introduces two more zero elements in the matrix
\begin{equation}
\label{coin:c:prime}
{\cal C}'=\left(\begin{array}{rrr}
0 & 0 & C_{13} \\
C_{21} & C_{22} & 0 \\
C_{31} & C_{32} & 0
\end{array}\right) \,,
\end{equation}
or
\begin{equation}
{\cal C}''=\left(\begin{array}{rrr}
0 & C_{12} & C_{13} \\
0 & C_{22} & C_{23} \\
C_{31} & 0 & 0
\end{array}\right) \,.
\end{equation}
The charateristic polynomial of $\tilde{U}(\mathbf{k})$ with coins
${\cal C}'$ or ${\cal C}''$ only depends on $\mathbf{e}_2$, thus the
eigenvalues depend only on $\mathbf{e}_2$. Consequently, the walk
propagates only on the direction linked to momentum $\mathbf{e}_2
\cdot \mathbf{k}$, namely $k_1$. In the direction $k_2$ the walk
decays exponentially, leading to a quasi $1$-dimensional walk.

It is easy to generalize the method shown above to see when
the element $C_{ii}$ is not zero and the other elements in
the diagonal of the coin operator are zero, then the walk is quasi
$1$-dimensional propagating only in the direction $\mathbf{e}_i$.
Note that if we introduce more zeros in the off-diagonals of ${\cal
C}'$ or ${\cal C}''$ then the matrices will be permutation matrices.
These permutation coins lead to trivial dynamics with no recurrence.

If the number of the zero elements on the diagonal of the
coin operator are less than two only isolated saddle points.
Hence quantum walks with this class of coins are transient.

The last interesting case is when all three diagonal elements of the
coin operator are zero. In this case, the equation
\ref{derived:saddle:line} has a solution for each ${\mathbf k_0}$,
i.e. $\omega_1(\mathbf{k})$ is constant and consequently the walk
will be localizing. Nevertheless, the coin operator is a permutation
matrix. We encounter here a trivial dynamics consisting in a mere
relabeling of the position states, in every third step the
completely relocalized initial state appears at the origin
independent of the initial coin state. Thus we obtain an
important result: there are no non-trivial localizing coin matrices
on a triangular lattice. 
%%%%%%%%%%%%%%%%%%%%%%%%

\begin{figure}
\includegraphics[width=0.4\textwidth]{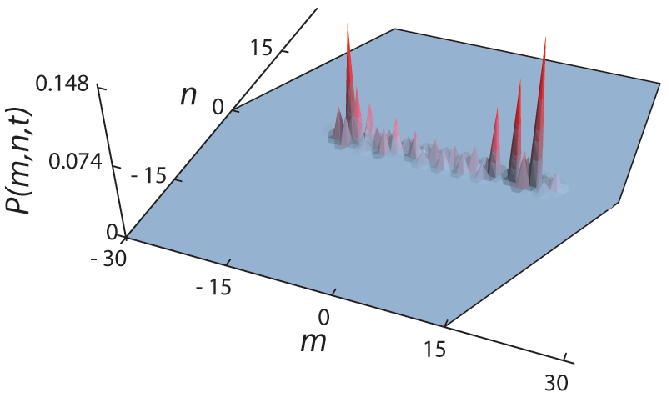}\vspace{12pt}
\includegraphics[width=0.4\textwidth]{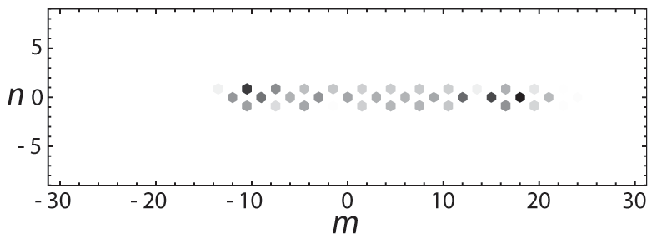}\vspace{12pt}
\includegraphics[width=0.4\textwidth]{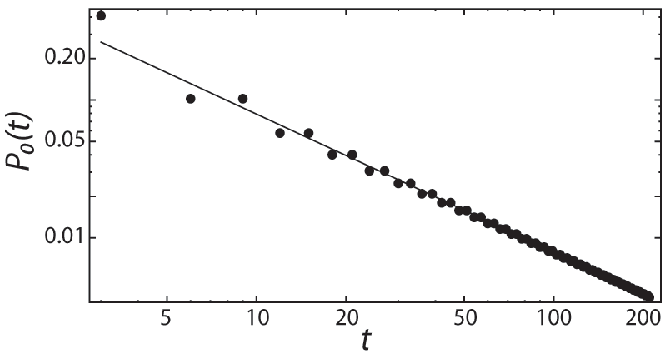}
\caption{(Color online) Probability distribution of the  walk driven by ${\cal
C}_{Rec}$  and the time dependence of the probability at the origin
for the initial state $\frac{1}{\sqrt{3}}(1,1,1)^T$. In the upper
plot, we show the probability distribution after 30 steps. The
center plot shows the quasi $1$-dimensional behavior of the same
probability distribution on a density plot with graylevel on the log
scale. The lower plot shows the $t^{-1}$ power law scaling of the
probability at the origin on a loglog scale. The straight line
corresponds to $t^{-1}$.} \label{fig:ehad}
\end{figure}

Illustrating the results above we constructed a coin operator which belongs to the class ${\cal C}'$

\begin{equation}
 \label{coin:ehadamard}
{\cal C}_{Rec} =  \frac{1}{\sqrt{2}}\left(
\begin{array}{rrr}
 0 &  0 & \sqrt{2} \\
 1 & 1 & 0 \\
 1 & -1 & 0
\end{array}
\right) \,.
\end{equation}

Let us analyze the behavior of a walk driven by  ${\cal C}_{Rec}$ observing the eigenenergies given by the characteristic polynomial
\begin{eqnarray}
\nonumber \Theta(\omega,\mathbf{k}) & = & \det\left(\tilde{U}(\mathbf{k}) - e^{i\omega} I\right) = \\
\nonumber & = & \cos\left( k_1 - \frac{\omega}{2}\right) - \sqrt{2} \cos\left( \frac{3 \omega}{2}\right) = 0\,.\\
\end{eqnarray}
The characteristic polynomial is independent of $k_2$, thus the
eigenenergies are independent of $k_2$, too. Hence the walk is quasi
$1$-dimensional (See \fig{fig:ehad}). Moreover, the double integral
in equation (\ref{psi:0}) simplifies to a single integral, leading
us back to the standard one-dimensional quantum walk
\cite{stefanak_prl_2008}. The decay rate of the probability at the
origin is $t^{-1}$ and the walk is recurrent,  as seen in
\fig{fig:ehad}.

%%%%%%%%%%%%%%%%%%%%%%%%%%%%%%%%%%%%%%%%%%%%%%%%%%%%%%%%%%%%%%%%%%%%%%%%%%%%%%%%%

\section{Summary}
\label{secsummary}

The analysis of three-state quantum walks on a triangular lattice
clearly demonstrates that neither the dimension of the underlying
lattice nor the dimension of the coin space in itself can determine
whether localization or recurrence can occur. Although we could
construct an example where the walk was recurrent, we have also
proven that only the coins from the special subclass leading to
quasi-onedimensional propagation allow for recurrence. We have
calculated the time exponents of the probability decay at the origin
for the Grover walk resulting in transience. This behavior is in
sharp contrast with the recurrence properties of the Grover walk
on a regular square lattices which traps the walker at
the origin with finite probability, also known as localization.

The triangular lattice is one of the few solvable models
allowing to decide about the recurrence properties of the quantum
walk. It is certainly interesting to analyze other higher
dimensional quantum walks from this perspective. The presented work
is just the fist step in classifying quantum walks using the concept
of recurrence and a number of interesting effects can be expected
for related models.

%%%%%%%%%%%%%%%%%%%%%%%%%%%%%%%%%%%%%%%%%%%%%%%%%%%%%%%%%%%%%%%%%%%%%%%%%%%%%%%

\begin{acknowledgments}

TK would like to thank Gy\"orgy K\'ali and Misha Titov for
interesting discussions. The financial support by MSM 6840770039,
M\v SMT LC 06002 and the Czech-Hungarian cooperation Project No. (MEB041011,CZ-11/2009) is gratefully acknowledged.

\end{acknowledgments}

%%%%%%%%%%%%%%%%%%%%%%%%%%%%%%%%%%%%%%%%%%%%%%%%%%%%%%%%%%%%%%%%%%%%%%%%%%%%%%%

\appendix

\section{The classical 3-way walk on $2$-dimensional Cartesian lattice}
\label{app:a}

The classical 3-way walk is strongly connected to the generalized Pascal's triangle, the so called Pascal's pyramid \cite{harris}, in a similar way as
the regular $1$-dimensional classical random walk on a line connects to Pascal's triangle. The central element in the $t$th Pascal's pyramid is given by
\begin{equation}
C_0(t) \equiv \frac{t!}{(t/3)!^3}\,.
\end{equation}
This expression is only valid when $t$ is dividable by 3 (as the walker could possible return to the origin at every 3rd step). We normalize the central element, to get the probability of the walker returning
\begin{equation}
p_0(t) \equiv \frac{1}{3^t} C_0(t)=\frac{1}{3^t}\frac{t!}{(t/3)!^3}\,.
\end{equation}

We approximate the return probability $p_0(t)$ with Stirling's approximation
\begin{equation}
p_0(t)=\frac{1}{3^t}\frac{t!}{(t/3)!^3} \approx \frac{3 \sqrt{3}}{2 \pi t}\,.
\end{equation}
The recurrence properties of the classical random walks are represented by the classical P\'olya number \cite{revesz}
\begin{equation}
\mathcal{P}=1 - \frac{1}{S}\,,
\end{equation}
where
\begin{equation}
S=\sum_{t=0}^\infty p_0(t)\,.
\end{equation}
In our case $p_0(t)$ is proportional to $t^{-1}$, hence the series $S$ diverges, the classical P\'olya number ($\mathcal{P}$) equals $1$, therefore the classical 3-way random walk is recurrent.

%%%%%%%%%%%%%%%%%%%%%%%%%%%%%%%%%%%%%%%%%%%%%%%%%%%%%%%%%%%%%%%%%%%%%%%%%%%%%%%

\end{document}